\documentclass[11pt]{llncs}

\usepackage{graphicx}
\usepackage{hyperref}
\usepackage{wrapfig}
\usepackage{xcolor}

\newcommand{\paragraphb}[1]{\vspace{0.03in}\noindent{\bf #1} }

\begin{document}
\title{Security Landscape for Robotics}
\author{Ryan Shah}
\institute{The University of Strathclyde}
\maketitle

\begin{abstract}
In this paper, the current state of security in robotics is described to be in need of review. When we consider safety mechanisms implemented in an Internet-connected robot, the requirement of safety becomes a crucial security requirement. Upon review of the current state of security in the field of robotics, four key requirements are in need of addressing: the supply chain for calibration, integrity and authenticity of commands (i.e. in teleoperation), physical-plane security and finally, secure, controlled logging and auditing.
\end{abstract}

\section{Introduction}
\label{sec:intro}

Robotics has taken a graceful leap into a wide range of application areas, including but not limited to autonomous vehicles, and surgical and industrial robotics. The use of robotics in these areas brings forth the potential to increase the efficiency of output, as well as the accuracy of operations. In the area of surgical robotics, a high level of accuracy and precision is a key factor, which, in the context of surgery, in some cases could mean the difference between life and death. Reuters published an article~\cite{reutersrobots} that states the shipments for robots increased by nearly 16\% from 2017, where shipments increased in every sector other than the automotive industry. An increase in shipments of robots and their corresponding components, ultimately corresponds to an increase in robotic installations, each of which are becoming Internet-connected. Whilst robots traditionally pose many safety challenges, becoming Internet-connected exposes the robots to cyber attacks which can result in safety-critical events, and thus safety problems now become a security concern.

\section{Background}
\label{sec:background}

\subsection{Surgical Robotics}

In the case of surgical robotics, there are robots which are configured based on pre-planned operations, such as bone-milling robots like Robodoc~\cite{taylor1999computer}, as well as robot systems which are directly teleoperated by the surgeon, such as RAVEN II~\cite{hannaford2013raven} and the da Vinci surgical system~\cite{tewari2002technique}.

Preoperative planning robots, like Robodoc, consist of a planning computer workstation and a robotic arm with an instrument, such as a high-speed milling device, as an end effector~\cite{bargar1998primary}. Before these robots were introduced, there was a need to address the inability to optimally fit implants into the bone of patients, as there are many factors which can impact the fitting of implants, including accurately sizing the implant and the precise positioning of the implant within the bone. As a result, preoperative planning robots designed for bone milling were introduced, which have shown to be feasible and provide successful surgery for implants, with a reduction in the number of post-surgery complications such as rejection of the implants and bone fractures. The precision and accuracy that these robots provide, allows for better bone milling at optimal speeds and locations which, due to a computer-operated preoperative planning system, is suitable for a wider variety of implant-related surgeries.

Teleoperated surgical robotic systems, such as da Vinci, provides medical professional with robot-assistance in procedures, to improve surgical performance. The da Vinci robot is a widely used surgical robot system, which is a quadbrachical (four-arm) robot, which is operated by a surgeon via a surgeon's console. The surgeon's console consists of finger controllers that translates human movements into instructions the robot can interpret. The RAVEN II surgical system consists of two surgical arms, but the motor chassis (base) allows two RAVEN II surgical robots to be mounted on either side of the surgery site~\cite{alemzadeh2016targeted}. Similarly, both of these robots are coupled with a surgical console, that provides feedback to the surgeon, such as by providing a 3D view of the surgical site.

Noticeably, with a comparison of both the RAVEN II and da Vinci robots, whose architectures span a majority of surgical robots, we can deduce that a key component across all surgical robotic systems is the surgeon's console. In the case of preoperative planning robots used in surgery, the console can be compared to the planning computer, where feedback to the surgeon's console such as radiography images (x rays, etc.) and robotic movements are used as input to the robot pre-surgery.

In a preoperative procedure, the surgical console is started and primed, and allowed to self-test itself, where it recognises its various components~\cite{tewari2002technique}. The cameras used in the surgical suite are black and white balanced and {\em calibrated} to a cross bar, as well as other factors such as the height are set to comfortable levels for the surgeon. Camera and arm movements are tested, and the system is left to stand by.

\subsection{Industrial Robotics}

Industrial robots accompany a large portion of installed robotics systems, which range from aiding in collection, production and packing of food products~\cite{botterill2017robot}, to the automotive industry.
An experimental security analysis of industrial robots~\cite{quarta2017experimental} shows that the majority of industrial robot architectures are extremely similar to surgical robot architectures. More specifically, industrial robot systems can be compared to teleoperated surgical robots, where the industrial robot is paired with a controller (pendant) that allow the robot to be both operated by human operators as well as autonomously. There is a main control system which links the controller, inter-connected components and the network in which the robot operates. Another comparison to surgical robotics, the control system is regarded as one of the most safety- and security-critical components. The components of the robot system communicate with each other using an internal robot network, and the robot system is connected to a dedicated organisational subnet or the Internet via a service network; similar to the da Vinci surgical system.
Traditional industrial robots are paired with a teach pendant that allows the robot to perform simple repetitive tasks, without any sensor feedback. The authors in~\cite{kruse2015sensor} describe a system to achieve coordinated task-based control for an industrial robot, which has a rotating torso and two seven degree-of-freedom arms to perform grasping tasks and object manipulation teleoperatively. Traditional robots are limited by a timing delay between the issuing of commands and the corresponding robotic action, but the authors show that the use of sensors allow the robot motion to be controlled with collision avoidance, object identification and locating. As well as this, using sensors and machine learning provide the robot with force-feedback control, where located objects can be securely held with enough force to prevent the object slipping, by estimating the weight and center of gravity of the object as well as its position. Another example of embedded sensors in an industrial robot is described in~\cite{botterill2017robot}, where the robot system uses cameras and sensors for constructing a 3D model used in automatic pruning of grape vines. The cameras of this robot are calibrated using standard methods and the robot-to-camera transform is computed by attaching a calibration target to the robot arm. To cut canes, a mill-end is attached to a 100W 24,000 RPM brushless DC motor on the arm. The 3D model is optimised using AI techniques and is used to update the robot's trajectory so that its six degree-of-freedom arm can reach the required vine cuts accurately.

\subsection{Autonomous and Unmanned Vehicles}

To a general audience, autonomous vehicles are primarily viewed as cars or other highway vehicles, yet this term is further classified. Autonomous vehicles can also be classified into civilian aircraft, trains, Unmanned Sea and Surface Vehicles (USVs), and Unmanned Aerial Vehicles (UAVs)~\cite{yaugdereli2015study}. Although the architectures of autonomous vehicles may differ among classes, they all utilise common components, such as cameras, sensors, GPS, etc. -- somewhat similar to surgical and industrial robots. These common components are heavily relied on, to aid with actions such as real-time manoeuvring decisions.

For autonomous cars, many in-built functions, such as enabling/disabling the brakes, are controlled by Electrical Control Units (ECUs), which are interconnected by the Controller Area Network (CAN) protocol~\cite{farsi1999overview}. Originally designed in the 1980s, the CAN bus is still the most widely used in-vehicle bus to date, due to its cost efficiency~\cite{groza2019efficient}. As well as this, a large proportion of autonomous systems can be directly addressed via cellular networks, and the Internet. In an autonomous vehicle, a combination of the ECUs in the vehicle are interconnected via the CAN bus, and thus we can portray autonomous vehicles to consist of a multi-agent system architecture, composed of one or more subsystems~\cite{yaugdereli2015study}. Similar to surgical robots, autonomous cars consist of one or more control modules, which may be connected to the CAN bus such that in can be interconnected with the ECUs in the vehicle~\cite{zong2018architecture}.

\section{Discussion of Robot Security}
\label{sec:discuss}

Upon exploring the problem space associated with security in three primary aspects of robotics, we identified key challenges that need to be addressed.

\subsection{Cyber Domain Challenges}

Attacks in the cyber domain are ever so present, when we connect robot systems to the Internet or even an organisation's network. Alemzadeh et al.~\cite{alemzadeh2016targeted} state that cyber domain attacks include the modification of control commands, without needing to modify the control flow or induce changes in the performance of the target program. The authors present attacks that exploit the time of check-time of use (TOCTOU) vulnerability, between the safety checks on the commands the actual execution of commands. As a result of attacks in the cyber domain, potentially catastrophic consequences arise in the physical domain, such as abrupt/jerked axial movements, which can lead to a reduction in accuracy and precision. Results like these are extremely hard to notice, unless closely watched by the operator, and it is hard to distinguish whether the result was due to a failure in the robot system, or if it was due to a human-induced failure. The challenges here include ensuring the integrity of commands whilst {\em in-flight}, maintaining the accuracy and precision of robot output, and providing tamper-resistant logs to maintain a trail of evidence in the event of failures or anomalous behaviour.

In surgical robotics systems, the key critical component is its electronic control system~\cite{alemzadeh2016targeted}, which both receives commands form the surgeon's console and translates these into robotic actions, whilst providing video feedback to the surgeon's console. Bonaci et al.~\cite{bonaci2015make} describe attacks between the surgeon's console and the robot itself, which they classify into three categories: {\em intention modification}, {\em intention manipulation}, and {\em hijacking attacks}. Respectively, these attacks include impacting intended actions of the surgeon, modifying feedback messages from the robot to the surgeon's console, and having the robot ignore the intended actions of the surgeon and ultimately perform a completely different action. Attacks such as these further suggest that providing tamper-resistant messages is important, and because commands are sent in real-time and executed as soon as they arrive, the integrity of commands must be maintained in-flight. On top of this, the communication channels between components in the robot system must not be tampered with, in the sense that messages within the channel shall not be altered, nor can unauthorised messages be sent and accepted within the channel.

In the case for industrial robots, with the architectures very similar to those of surgical robots, the critical component is also the control system. Industrial robots tend to have components interconnected and communicate with each other through an internal robot network, as well as a service network where the robot is connected to a dedicated organisational subnet and possibly to the Internet~\cite{quarta2017experimental}. What is challenging, is to secure robots that are connected to the public Internet, where adversaries can target the robot (or multiple) from anywhere. On top of this, even when a robot is connected to the organisational subnet, there is still a risk of local adversaries that may look to attack the robot in the cyber domain. Demarinis et al.~\cite{demarinis2018scanning} present results from scanning the entire IPv4 address space of the Internet for instances of the Robot Operating System (ROS), which is a widely used robotics platform for research, and found that a number of hosts that support ROS are exposed to the public Internet. They identified that some robots are connected to simulators, but a number appear to be legitimate robots which are capable of being teleoperated, such that they can be remotely moved in ways that could potentially pose risks to both the robot and its surrounding environment, including human operators. An open ROS master was shown to indicate a robot whose sensors can be remotely accessed, and who's actuators can be remotely controlled. From a single scan, 15 instances of the operating system were identified, with 11 of them located in cloud service provider networks. Multiple different sensors were also discovered to be in use by these openly accessible robots, which included laser range finders, cameras, GPS devices and barometric pressure sensors. Ultimately, the authors suggest that, at a minimum, the network should be monitored to detect exposure. The main challenge here is to only allow entities to teleoperate upon authorisation, whilst either securing or isolating the robot from the public Internet.

While safety is becoming of increasing importance, the implementation of safety mechanisms in software establishes higher security requirements. Focusing attacks on these safety mechanisms can subsequently impact the physical domain. Impacts on the operations of the robot itself can occur, but also indirect impacts on its operators and surrounding environment may be ever so present. An indicative example of this relates back to an attack which can cause abrupt, unjust movements in the robot arms~\cite{alemzadeh2016targeted}. In this example, the authors uncover an interesting challenge with the E-stop (emergency stop) software mechanism. This mechanism can also be hard-wired, but their experiment involved a software e-stop mechanism. Assuming that a robot has been compromised, installing a safety mechanism which can impose a system-wide halt would be prone to problems. For example, if a rival manufacturing company was to attack the e-stop software mechanism, they could interrupt the entire manufacturing process of their competitors. This not only leads back to the challenges of securing the robot itself, but also providing a tamper-resistant log of evidence, in the event that this may take place; such that unintended actions can be traced back to the origin of the action.

% Cyber attacks on autonomous vehicles

\subsection{Physical domain challenges}

Aside from challenges that arise in the cyber domain, we must also consider attacks in the physical domain; as the connected robots consist of both cyber and physical components which cooperate to enable smooth operation in the real world. 

First, we must consider physical compromise. The availability of these robots is regarded as a property of uttermost importance, which is unfortunately also the most easily influenced. For example, an adversary who targets an autonomous vehicle with a jamming attack, focusing on the optical channels, could induce an emergency stop resulting in a denial-of-service (DoS) attack. Aside from jamming attacks, even tampering with a sensor or replacing it with an unauthorised sensor could provide images with unintended perturbations or of low quality, which can result in poor misclassification of objects in a robots perceived environment.
Therefore, we must consider the challenges of not only mitigating the effects of jamming attacks, but also avoid unauthorised equipment to be added (or in replacement of another) to robots. Furthermore, it is also a challenge to ensure only authenticated components are in place to ensure that the robot and its components behave and perform as intended.

As well as safety mechanisms implemented in software, we must also discuss challenges posed by safety mechanisms in the physical domain.
Industrial robots have been discussed for use in construction~\cite{salmi2018human}, which poses many interesting challenges, as construction
work invokes higher requirements for flexibility and adaptability, compared
to traditional industrial robot solutions. The working environment in construction is ever changing, with humans potentially cooperating aside the robot itself, and materials coming in and out of the robots perceived environment. Compared to typical industrial environments, the accuracy of components and buildings vary and are relatively lower, and thus must be compensated for. Salmi et al.~\cite{salmi2018human} note that using sensors for 3D vision and force control allow robots to better manage complex environments, as well as adapt to product variation and adapt process parameters to varying requirements. The use of sensors as a safety mechanism allows for flexible fenceless safety systems, which is ideal in environments where humans and robots closely collaborate. However, risks such as collision, compression and sparks, among others, can occur at different stages of robot operation (i.e. boot, clearing of disturbances, etc.) even with the use of sensors. Sensors can be expensive, but~\cite{kruse2015sensor} shows that off-the-shelf sensor components and software can be drawn upon, which can greatly reduce costs. Using cheaper sensors, albeit a cheaper option, does not guarantee a long component-life. Furthermore, the calibration of these sensors can pose a threat to the accuracy and precision of these sensors, and ultimately its effectiveness as a safety mechanism.

\subsection{Secure calibration}

% Calibration background

The calibration of any equipment, including sensors used in AVs or even the instruments used for surgical robotics, is performed to ensure that the output of the equipment are of high quality, precision and accuracy, and low uncertainty. Moreover, the definition of equipment goes further to also cover cables and resistors that are used in the equipment itself. To enable this, they are calibrated against a higher standard of accuracy, to identify the margin of errors in output. Calibration is also performed to meet quality audit requirements, and ensure reference designs, subsystems and integrated systems perform as intended. After calibration is performed, a calibration certificate is produced as an output, which outlines the measurement uncertainty, error margins and precision under certain environmental conditions (dependent on calibration), amongst others. A piece of equipment's calibration and uncertainty should be traceable to corresponding SI units, to a standard or reference method~\cite{de2000calibration} -- producing a chain of equipment and associated calibration certificates, leading to the SI units. We can define a simple traceability chain for a sensor, where the sensor is calibrated at one or more intermediate calibration facility, against one or more calibration units. These calibration units are themselves calibrated against some master calibration units, and this continues until we reach calibration performed at National Measurement Institutes (NMIs), which use calibration units that are calibrated against the SI units; providing the highest levels of accuracy and lowest levels of uncertainty. As we move from the NMIs at the top of the chain, the accuracy decreases and uncertainty increases, but equipment at each level should have a set uncertainty; which should be maintained.

The challenges associated with robot equipment are ensuring that all its components are correctly calibrated, whilst maintaining the integrity of calibration and certificates. Aside from integrity, we must also ensure that for some equipment, i.e. a sensor, its traceability chain is unbroken. Specifically, if one master unit in its traceability chain is out of calibration (needing recalibration) or was calibrated incorrectly, all subsequent equipment/units downwards from that point should effectively be recalibrated.

As well as challenges associated with implemented safety mechanisms of robots, Quarta et al.~\cite{quarta2017experimental} note a key challenge associated with calibration parameters. These calibration parameters are an essential construct for determining factors such as axial positions of the arms, which are used for compensation for known measurement errors. By manipulating the calibration parameters, they demonstrated induced effects on the robot's servo motor, causing it to move erratically. This was because the true error in the measurement signal was in fact different to that which the controller knew. Calibration parameters for all components in a robot system may be held by its controller or main computer. If this is the case, then any potential compromise of the controller or main computer can subsequently violate the integrity of calibration parameters, thus leading to potentially anomalous output of sensors or even unintended actions even if the robot was operated normally.

\subsection{Secure boot}

Importantly, calibration, configuration and component connections are all established through the boot process. By tampering with only the boot process, an adversary could modify the calibration and configuration parameters known by the controller, and potentially force extreme erratic movements or unsafe robot starting positions. Aside from tampering with the boot process, incorrect calibration, faulty connections, etc. could result in potentially catastrophic consequences. Therefore, assuming a secure boot, the boot process should ensure factors such as authentic component connections and valid calibration, before the robot can operate safely.

\subsection{Summary of challenges}

Overall, we have established a number of challenges that arise in security of robotics, from cyber and physical domain challenges, to further refined challenges in the calibration of interconnected components such as sensors and even cables used to connect components.

\begin{enumerate}
	\item{Ensuring the integrity and authenticity of commands, even whilst in-flight.}
	\item{Maintaining tamper-resistant logs and trails of evidence, for auditing and locating sources of failures and anomalous/unsafe behaviour.}
	\item{Securing communication channels from tampering commands, adversarial input, and eavesdropping.}
	\item{Securing the teleoperability of robots -- only authorised entities can teleoperate, even if a robot is connected to the public Internet or an organisation's subnet.}
	\item{Mitigation of physical plane attacks, such as jamming and signal saturation attacks.}
	\item{Ensuring only authentic and authorised equipment is in place, to ensure no physical tampering occurs. This includes sensors, and actuators, while going even further to include components such as power and ethernet cables.}
	\item{Maintaining valid calibration for all components in a robot system, to ensure high accuracy and precision, and low uncertainty (low margin of error from calibration).}
	\item{Maintaining the integrity of calibration certificates and stored calibration/configuration.}
	\item{Securing the boot process (startup).}
\end{enumerate}

\section{The Requirement for a Progressive Shift in Robot Security}

Upon discussion of the challenges associated with robot security, we propose a number of research questions which are to be further explored.

\vspace{0.4cm}

\paragraphb{Supply chain for calibration:} First, {\em how can we secure the supply chain, before a robot is produced?} The majority of components that make up a robot system have a calibration a chain, which means that they have a calibration certificate detailing accuracy, precision and margins of error. Thus, it is a requirement to ensure that the calibration of these components is up-to-date and performed correctly and to international standards~\cite{isocalibration}. Furthermore, not only should the calibration of a component be valid, the calibration of all calibration units that supersedes must also be valid such that the traceability chain remains unbroken. If part of this chain is broken, all subsequent members of the chain must be recalibrated. In a real-world scenario, assuming a sensor was out-of-calibration, the robot may perform less efficiently or incorrectly classify sensed objects; and such the question of whether the robot should still be in operation becomes of question. This can be taken a step further, and the proper calibration of a robot can be determined in a secure boot process which ultimately determines whether a robot can start or not.

\vspace{0.4cm}

\paragraphb{Integrity and authenticity of commands (i.e. from teleoperation), even whilst in-flight:} Bonaci et al.~\cite{bonaci2015make} describe an attack which impacts a surgeon's intended actions and modifying in-flight packets. Doing so could result in the robot assuming a surgeon's commands are valid and ultimately perform a consequential action targeting a patient. As a solution to attacks against the integrity and authenticity of commands, the authors propose a number of challenges to be addressed. First, teleoperated robots execute commands as soon as they are received, and thus if an adversary was to flood the robot with packets, it may start behaving abnormally. Not only would this pose a risk to a patient, but also to the robot and its environment. As a minimum requirement, they propose that assurance of legitimate sources of commands should be maintained and to limit a robot's processing rate to a specific amount to protect the robot and its environment from harm.

\vspace{0.4cm}

\paragraphb{Physical plane security:} Assuming that we have a secure supply chain, once the robot is produced and ready to be used, {\em how can we ensure it is physically secure?} We must first ensure that robots are able to maintain relatively normal operation even under adversarial threats. For example, if jamming or signal saturation attacks are in place to disrupt a driverless vehicle, we must mitigate the effects of these attacks as best as possible, to enable the vehicle to continue driving regularly. On top of this, we must ensure that all interconnected components of a robot system are authentic and allowed to be there. Specifically, we must ensure that sensors installed during production have not been changed unless the change was authorised and logged securely. This would help prevent faulty components being used, as well as preventing an adversary from implementing a modified component for their benefit. For example, if a UAV was captured in a battlefield scenario, an attacker may modify its GPS component so it provides incorrect information to the owner of the UAV. Furthermore, even cables can be used maliciously. USB cables can seemingly be legitimate, yet an attacker can modify the USB cables to inject malware, as demonstrated by Nohl and Lell~\cite{badusb}. Therefore, it is a requirement to ensure that all components are authentic and that even component connections should be authorised. For example, a challenge should be made to verify that a cable is allowed to be attached to an Ethernet port. Finally, robots should also be securely maintained. This can pose a difficult requirement to address, as some robots operate remotely or are always running, and therefore it is difficult to perform maintenance on a live robot. Can we patch the robot, even whilst it is in operation, without disrupting its operability? Can we, securely and safely, lock or shutdown the robot in the event of compromise or abnormal behaviour?

\vspace{0.4cm}

\paragraphb{Secure, controlled logging and auditing:} The need for tamper-resistant logs is a valuable requirement, which can provide a trail of evidence in catastrophic events, for example to locate the source of potential problems. As well as this, logs are also used for auditing - which is an essential event that occurs regularly within the calibration infrastructure. There has been several research papers focusing on append-only~\cite{yavuz2012efficient}, tamper-evident~\cite{crosby2009efficient} and tamper-resistant logs~\cite{ma2009new} and secure audit logging, which primarily focus on publicly verifiable audit logs~\cite{yavuz2009baf,holt2006logcrypt,waqas2013fault,xie2012tiered} for distributed systems. However, there is little work done on privacy-preserving audit logging. At the least, work on privacy-preserving audit logging focuses on inter-domain audit logging~\cite{lee2006privacy,pulls2013distributed,shah2008privacy}, but in a robotics environment, with multiple levels each of different security requirements (i.e. confidentiality and integrity), it is important to establish a framework for cross-domain, multi-level, privacy-preserving audit logging.

% --- end --- %

\bibliographystyle{unsrt}
\bibliography{references}

\end{document}